\documentclass{article}

\usepackage[numbers]{natbib}
\usepackage[shortlabels]{enumitem}
\usepackage[vcentering,dvips]{geometry}
\usepackage[usenames, dvipsnames]{color}
\usepackage{amsmath,amssymb,amsfonts,amsthm}
\usepackage{float}
\usepackage{enumerate}
\usepackage{graphicx}
\usepackage[
hidelinks,
pageanchor=true,
plainpages=false,
pdfpagelabels,
bookmarks,
bookmarksnumbered
]{hyperref}
\usepackage{mathtools,bbm}
\usepackage{moreverb}
\usepackage{setspace}
\usepackage{url}
\usepackage{mathrsfs}
\usepackage{dsfont}

\usepackage{algorithm}
\usepackage{algpseudocode}

\newlength\figureheight
\newlength\figurewidth
\DeclareMathOperator*{\argmin}{arg\,min}
\DeclareMathOperator*{\argmax}{arg\,max}

\newcommand{\real}{ {\mathbb{R}} }

\newcommand{\setsystemfont}[1]{{\mathcal{#1}}}
\newcommand{\ground}{\Omega}
\newcommand{\independence}{\setsystemfont{I}}

\newcommand{\curvature}{\nu}
\newcommand{\sol}{A^\mathrm{sol}}
\newcommand{\opt}{A^\star}
\newcommand{\optremoval}{B^\star}
\newcommand{\remain}{R}
\newcommand{\remainopt}{R^{\mathrm{a}\star}}

\newtheorem{theorem}{Theorem}
\newtheorem{definition}{Definition}
\newtheorem{problem}{Problem}

\title{A Simple Bound for Resilient Submodular Maximization with Curvature}
\author{Micah Corah%
\thanks{This work was completed while Micah was a student at Carnegie Mellon
University. Email: \texttt{micah.d.corah@jpl.nasa.gov}}}
\date{}

\begin{document}
\maketitle

\section{Introduction}

Resilient submodular maximization refers to the combinatorial problems studied
by \citet{nemhauser1978} and \citet{fisher1978} and asks how to maximize an
objective given a number of adversarial removals.
For example, one application of this problem is multi-robot sensor planning with
adversarial attacks~\citep{zhou2018ral,schlotfeldt2018iros}.
However, more general applications of submodular maximization are also
relevant.
\citet{tzoumas2018arxiv} obtain near-optimal solutions to this problem by taking
advantage of a property called \emph{curvature} to produce a mechanism which
makes certain \emph{bait} elements interchangeable with other elements of the
solution that are produced via typical \emph{greedy}
means~\citep{fisher1978,nemhauser1978}.
This document demonstrates that---at least in theory---applying the method for
selection of bait elements to the entire solution can improve that guarantee on
solution quality.

Specifically, this document provides a brief description and proof for an
algorithm that improves upon bounds for resilient submodular maximization with
curvature by \citet[(6), (7)]{tzoumas2018arxiv}.
The approach builds on work by \citet{zhou2020icra} and takes aspects of that
analysis to a natural conclusion.\footnote{
  This document arises, in part, from discussions with Vasileios Tzoumas and
  Lifeng Zhou.
  Additionally, another version of this analysis appears
  in~\citep{zhou2020arxiv}.
}

The proposed algorithm maximizes objective values for
individual solution elements on their own i.e. $f(x)$ for $x\in\ground$
subject to some constraint where $f$ is a submodular, monotonic, normalized set
function and $\ground$ is the ground set.\footnote{%
  For brevity we omit some definitions.
  Please refer to existing works on submodular
  maximization~\citep{nemhauser1978,fisher1978}
  for more detailed discussion.
}
This contrasts with typical methods which maximize marginal
gains of the form $f(x) - f(X)$ where $X\subseteq\ground$ is typically a partial
solution.
Intuitively, the approach that this document describes will perform well because
the curvature of $f$ prevents objective values for sets from deviating far
from the sum of values for solution elements considered individually.

Curvature is defined as follows:
\begin{definition}[Curvature]
  Consider a monotonic and submodular set function $f$.
  The curvature of $f$ is
  \begin{align}
    \curvature = 1 - \min_{x\in\ground}
    \frac{f(\ground) - f(\ground \setminus x)}{f(x)}.
    \label{eq:curvature}
  \end{align}
\end{definition}

Additionally, the resilient submodular maximization problem seeks to maximize
$f$ given a number $\alpha$ of adversarial removals:\footnote{
  In this document, the removals are subject only to a cardinality
  constraint (forming a uniform matroid).
  \citet{tzoumas2018arxiv} also discuss a limited class of problems where
  the removals are subject to a partition matroid constraint.
  Although we do not present results for this class of problems directly,
  similar results can be obtained and produce same bound.
}

\begin{problem}[Resilient submodular maximization]
  \label{problem:resilient}
  Consider a submodular, monotonic, normalized set function $f : 2^\ground
  \rightarrow \real$ where $\ground$ is the ground set and
  $(\independence, \ground)$ is a matroid.
  The goal is to maximize $f$ subject to $\alpha$ worst-case removals.
  That is,
  \begin{align}
    \max_{A \in \independence}
    \
    \min_{B \subseteq A,\ |B| \leq \alpha}
    \
    f(A \setminus B).
    \label{eq:resilient}
  \end{align}
\end{problem}

\noindent
Selecting solution elements for value on their own $f(x)$ will admit exchange of
individual solution elements (which solve maximization subproblems exactly) for
elements of the optimal solution to obtain an approximation guarantee of
$1-\curvature$.
However, applying methods for greedy submodular
maximization~\citep[(41)--(43)]{tzoumas2018arxiv} incurs a penalty in the form
of the suboptimality of the greedy procedure for an approximation factor of
$\frac{1-\curvature}{1+\curvature}$.

In the following, we assume familiarity with background material (matroids, set
functions, monotonicity, and submodularity) although readers may refer to
related works for background material~\citep{tzoumas2018arxiv}.

\section{Approach}

Algorithm~\ref{alg:myopic} produces feasible, suboptimal solutions $\sol$ to
Prob.~\ref{problem:resilient} by repeatedly maximizing objective values for
individual solution elements until it obtains a complete solution.

Additionally, if $(\ground, \independence)$ is a partition matroid, maximization
steps can run over individual blocks of the partition as in local greedy
methods~\citep{fisher1978,goundan2007,singh2009} and in parallel as well.
Likewise, maximization steps can be executed in a  distributed manner such as
for multi-robot applications~\citep{zhou2020icra}.

\begin{algorithm}[h!]
  \caption{Myopic maximization}\label{alg:myopic}
  \begin{minipage}{\linewidth}
    \begin{algorithmic}[1]
      \State $\sol \gets \emptyset$
      \While{$|\sol| < \mathrm{rank}(\independence)$}
        \State $a \gets
          \argmax_{
            a \in \ground \setminus \sol,\
            \sol \cup \{a\} \in \independence
          }
          f(\{a\})$
          \label{line:maximization}
        \State $\sol \gets \sol \cup \{a\}$
      \EndWhile
      \State \Return $\sol$
    \end{algorithmic}
  \end{minipage}
\end{algorithm}

\section{Analysis}

Let us begin with a few definitions.
Let $\opt$ be an optimal solution to Prob.~\ref{problem:resilient}, and
let $\optremoval(A)$ be an optimal removal so that for any
$A\in\independence$, then
\begin{align}
  \optremoval(A) \in \argmin_{B \subseteq A,\ |B| \leq \alpha}
  f(A \setminus B).
  \label{eq:optimal_removal}
\end{align}
For brevity, the \emph{remaining} solution elements after removing
$\optremoval(A)$
are denoted
$\remain(A) = A \setminus \optremoval(A)$

We can now state the main result:

\begin{theorem}[Approximation performance]
  Algorithm~\ref{alg:myopic} outputs feasible solutions to
  Prob.~\ref{problem:resilient} that satisfy
  \begin{align}
    f(\remain(\sol)) \geq
    (1 - \curvature)
    f(\remain(\opt)),
  \end{align}
  observing that $f(\remain(\opt))$ is the optimal objective value for
  \eqref{eq:resilient}.
\end{theorem}
\begin{proof}
  Due to curvature~\eqref{eq:curvature}, submodularity, and monotonicity,
  objective values over remaining elements are lower bounded in terms of sums
  over individual elements as
  \begin{align}
    f(\remain(\sol))
    \geq
    (1-\curvature) \sum_{a \in \remain(\sol)} f(a),
    \label{eq:proof_curvature}
  \end{align}
  and this statement follows from~\citep[Lemma~2]{tzoumas2018arxiv}.

  Now, to account for the partition matroid, \citep[Theorem 1]{brualdi1969}
  implies existence of a bijection
  $\pi : \sol \rightarrow \opt$ between $\sol$ and $\opt$
  (which are each bases of $\independence$)
  that enables exchange of elements between the two solutions
  so that for any $a\in\sol$ then
  $\sol \setminus \{a\} \cup \{\pi(a)\} \in \independence$.
  Further, $\pi$ may be defined as the identity on the intersection $\sol \cap
  \opt$.\footnote{%
    For an example, please refer to discussion by \citet{filmus2014}.
  }
  That is, either $a=\pi(a)$
  (the given solution element is also part of the optimal solution)
  or $\pi(a)\notin \sol$, and $\pi(a)$ was also a feasible
  solution to the maximization step at which $a$ was selected
  (Alg.~\ref{alg:myopic}, line~\ref{line:maximization}).
  Then, following from these two cases of the identity and feasibility during the
  maximization step,
  \begin{align}
    f(a) \geq f(\pi(a)), \quad \mbox{for all}\ a \in \sol.
    \label{eq:greedy_choice}
  \end{align}
  Substituting \eqref{eq:greedy_choice} into \eqref{eq:proof_curvature},
  we get
  \begin{align}
    f(\remain(\sol))
    \geq
    (1-\curvature)
    \sum_{a \in \remain(\sol)} f(\pi(a)).
    \label{eq:proof_exchange}
  \end{align}
  Then, define $\remainopt = \{\pi(a) \mid a \in \remain(\sol)\}$ as the subset
  of the optimal solution ($\remainopt \subseteq \opt$) with elements
  corresponding to the remaining elements of the our suboptimal solution.
  Applying submodularity to $\eqref{eq:proof_exchange}$ and writing the result
  in terms of $\remainopt$,
  \begin{align}
    f(\remain(\sol))
    \geq
    (1-\curvature)
    f(\remainopt).
    \label{eq:mapped_remainder}
  \end{align}
  Given that $|\opt \setminus \remainopt|=\alpha$, equation
  \eqref{eq:optimal_removal} implies that $f(\remainopt)
  \geq f(R(\opt))$.
  Then, substituting back into \eqref{eq:mapped_remainder},
  \begin{align}
    f(\remain(\sol))
    \geq
    (1-\curvature)
    f(R(\opt)).
  \end{align}
  This completes the proof.
\end{proof}

\section{Conclusion}

The algorithm which this document describes is trivial and likely not of much
interest for any applications.
Yet, the approximation guarantee improves on results by \citet{tzoumas2018arxiv}
which have since been applied in several recent
works~\citep{schlotfeldt2018iros,zhou2018ral,zhou2020icra}.
Moreover, application, in part, of a greedy algorithm~\citep{tzoumas2018arxiv}
is desirable and may produce superior performance in practice if the number of
removals is uncertain or if the removals are not actually adversarial.
Instead, these results may encourage further improvements
to suboptimality guarantees for Prob.~\ref{problem:resilient}
or study of more specialized resilient submodular maximization problems that
may exhibit different behaviors.

\section*{Acknowledgments}

I would like to thank Vasileios Tzoumas and Lifeng Zhou for looking over and
discussing this document.

\bibliographystyle{plainnat}
\bibliography{bibliography}

\end{document}